\documentclass[fleqn,useAMS,usenatbib]{mn2e}
\usepackage{graphicx}
\usepackage{amsmath}
\usepackage{amssymb}

\title[PDF of a 3D density field from 2D observations]{A method for reconstructing the PDF of a 3D turbulent density field from 2D observations}
\author[C. M. Brunt, C. Federrath, \& D. J. Price ]{Christopher M. Brunt$^{1}$\thanks{E-mail brunt@astro.ex.ac.uk}, Christoph Federrath$^{2,3}$, \& Daniel J. Price$^{4}$\\
$^{1}$School of Physics, University of Exeter, Stocker Road, Exeter, UK\\
$^{2}$Zentrum f\"{u}r Astronomie der Universit\"{a}t Heidelberg, Institut f\"{u}r Theoretische Astrophysik, Albert-Ueberle-Str. 2, \\ D-69120 Heidelberg, Germany\\
$^{3}$Max-Planck-Institute for Astronomy, K\"{o}nigstuhl 17, D-69117 Heidelberg, Germany \\
$^{4}$Centre for Stellar and Planetary Astrophysics, School of Mathematical Sciences, Monash University, Clayton Vic 3168, Australia}
\begin{document}

\date{Accepted ; Received ; in original form }

\pagerange{\pageref{firstpage}--\pageref{lastpage}} \pubyear{2009}

\maketitle

\label{firstpage}

\begin{abstract}
We introduce a method for calculating the probability density function (PDF) of a turbulent density 
field in three dimensions using only information contained in the projected two-dimensional column density field. 
We test the method by applying it to numerical simulations of hydrodynamic and magnetohydrodynamic turbulence in molecular clouds.
To a good approximation, the PDF of log(normalised column density) is a compressed, shifted version of the PDF of
log(normalised density). The degree of compression can be determined observationally from the column density power spectrum,
under the assumption of statistical isotropy of the turbulence.
\end{abstract}


\section{Introduction}

The probability density function (PDF) of the density field in molecular clouds is
a key ingredient in most analytic models of star formation
(Padoan \& Nordlund 2002; Krumholz \& McKee 2005; Elmegreen 2008; 
Hennebelle \& Chabrier 2008; Padoan \& Nordlund 2009; Hennebelle \& Chabrier 2009).
Knowledge of the density PDF is required to determine the overall star formation
rate or efficiency. In some models, the density PDF is of central importance
in determining the emergent stellar initial mass function. The majority
of models assume a lognormal density PDF (V{\'a}zquez-Semadeni 1994), with the 
width of the PDF increasing with the Mach number of the turbulence 
(Padoan, Nordlund, \& Jones 1997; Passot \& V\'{a}zquez-Semadeni 1998;
Federrath, Klessen, \& Schmidt 2008).

Observational knowledge of density fields in molecular clouds (let alone the
PDF) is very limited. We do not have access to the density field in three 
dimensions (3D) but instead can only view the projected column density
field in two dimensions (2D). Use of molecular tracers of different critical
density can in principle yield some information, but even the most sophisticated
excitation analyses provide only a single ``density'' per line-of-sight whereas 
the transverse variations in ``density'' so measured must imply comparable, or
perhaps greater, fluctuations in density along the line-of-sight. A route
to the PDF could involve (for example) the measurement of mass exceeding
a range of critical densities from a suite of tracers. While the amount of
data involved here would likely be prohibitive for nearby clouds of large angular
extent, it can be usefully applied in an extragalactic context (e.g.
Krumholz \& Thompson 2007). 

Column density fields traced by extinction of background stars are perhaps
the most robust way of acquiring constraining data on the density PDF in
nearby clouds. 
Column density PDFs can be constructed (e.g. Cambresy 1999; Lombardi 2009;
Kainulainen et al 2009)
but the relation between these and the 3D density PDF is currently unknown.
Compression of the PDF due to line-of-sight averaging is expected, and
there are some indications that a lognormal density PDF will project into
a (less broad) lognormal column density PDF (Ostriker, Stone, \& Gammie 2001;
V{\'a}zquez-Semadeni \& Garc{\'i}a 2001, Federrath et al 2009). The degree of compression is
presently unknown, but recent work by Brunt, Federrath, \& Price (2010; hereafter BFP)
demonstrated how to calculate the normalised density variance in 3D from
information contained solely in the column density field. Comparison of
the measured normalised column density variance with the inferred normalised 
density variance can provide some information on the degree of compression.

In this paper, we introduce a method by which
the 3D density PDF can be constructed from measurements made on the column
density field alone. Measurements of the normalised column density variance,
the column density power spectrum, and the column density PDF are required,
and can be combined to construct an estimate of the 3D density PDF.

\section{The Method}

\subsection{Reconstructing the 3D Density Variance}

We define the normalised density field in 3D, $x_{3}$, as:
\begin{equation}
x_{3} = \rho / \rho_{0} ,
\label{eqno1}
\end{equation}
where $\rho$ is the density and $\rho_{0}$ is the mean density. Similarly,
in 2D, the normalised column density, $x_{2}$, is defined as:
\begin{equation}
x_{2} = N / N_{0} ,
\label{eqno2}
\end{equation}
where $N$ is the column density and $N_{0}$ is the mean column density.

The normalised density variance, $\sigma^{2}_{x_{3}}$ is given by:
\begin{equation}
\sigma^{2}_{x_{3}} = \langle x^{2}_{3} \rangle - \langle x_{3} \rangle^{2} ,
\label{eqno3}
\end{equation}
and the normalised column density variance, $\sigma^{2}_{x_{2}}$, is given by:
\begin{equation}
\sigma^{2}_{x_{2}} = \langle x^{2}_{2} \rangle - \langle x_{2} \rangle^{2} ,
\label{eqno4}
\end{equation}
where angle brackets denote averaging over the fields. 

While the field $x_{3}$ is observationally inaccessible, BFP showed that $\sigma^{2}_{x_{3}}$ 
can be estimated solely from measurements made on the field $x_{2}$.
Making use of Parseval's Theorem, BFP define the 2D-to-3D variance ratio, $R$, as:
\begin{equation}
R = \frac{ \left( \displaystyle\sum_{k_{x}=-\lambda/2+1}^{\lambda/2} \sum_{k_{y}=-\lambda/2+1}^{\lambda/2} \langle P_{x_{2}} \rangle (k)  \right) - P_{x_{2}}(0)}{ \left( \displaystyle\sum_{k_{x}=-\lambda/2+1}^{\lambda/2} \sum_{k_{y}=-\lambda/2+1}^{\lambda/2} \sum_{k_{z}=-\lambda/2+1}^{\lambda/2} \langle P_{x_{2}} \rangle (k) \right) - P_{x_{2}}(0)} ,
\label{eqno5}
\end{equation}
where  $\langle P_{x_{2}} \rangle (k)$ is the azimuthally averaged power spectrum of $x_{2}$,
$P_{x_{2}}(0)$ is the power spectrum of $x_{2}$ evaluated at $k = 0$, 
and $\lambda$ is the scale ratio of the field (i.e. the number of pixels along each axis, with the
requirement that the field is square). 

The inferred 3D density variance, $\sigma^{2}_{x_{3R}}$, can then be calculated via:
\begin{equation}
\sigma^{2}_{x_{3R}} = \sigma^{2}_{x_{2}} / R ,
\label{eqno6}
\end{equation}
and BFP show that:
\begin{equation}
\sigma^{2}_{x_{3R}} \approx \sigma^{2}_{x_{3}} 
\label{eqno7}
\end{equation}
to about 10\% accuracy as long as the field $x_{3}$ is statistically isotropic (see BFP for
a discussion of this requirement). 
Straightforward modifications of equation~(\ref{eqno5}) and equation~(\ref{eqno6})
to account for the effect of a telescope beam and to account for the effect of zero-padding (to
produce a square field and/or to reduce edge discontinuities in 
the power spectrum calculation) are given in BFP. 

It is important to recognise that the variance of a field calculated at finite resolution
is necessarily a lower limit to the variance that would be observed in the continuous
limit (i.e. arbitrarily high resolution). Brunt (2009) discusses the consequences of 
this when the BFP method is applied to
the Taurus Molecular Cloud, finding that $\sigma^{2}_{x_{3R}}$ may underestimate the
variance that would be obtained in the continuous limit by as much as a factor of 2. For
the numerical simulations used in this paper, there is no sub-resolution structure, and
the 3D variances are not subject to this problem. 

\subsection{Reconstructing the 3D Density PDF}

In the following analysis, we propose to make a transformation of the
field $x_{2}$ such that the normalised variance of the transformed field
is equal to $\sigma^{2}_{x_{3R}}$. Our conjecture is that the transformed
field has the same PDF as the 3D normalised density field $x_{3}$. Note that a 
suitable transformation can match the first two moments of the normalised transformed 
field to the first two moments of the normalised 3D density field (mean = unity,
variance = $\sigma^{2}_{x_{3R}}$~$\approx$~$\sigma^{2}_{x_{3}}$). We do not
have access to the higher order 3D moments, so must rely on experiment to
investigate the reliability of this procedure.

We begin by noting that a transformation $x_{2} \longrightarrow ax_{2}$, where $a$
is a constant, has no effect on the normalised variance due to the re-normalisation 
of $ax_{2}$. The simplest useful transformation is therefore $x_{2} \longrightarrow ax_{2}^{\xi}$
where $a$ and $\xi$ are constants. We therefore define the transformed field, $x_{3R}$, via:
\begin{equation}
x_{3R} = a x_{2}^{\xi} ,
\label{eqno8}
\end{equation}
where the re-normalising constant, $a$, is given by:
\begin{equation}
a = \langle x_{2}^{\xi} \rangle ^{-1} ,
\label{eqno9}
\end{equation}
and we have called the transformed field $x_{3R}$ (even though it is technically a 2D field)
since we intend to endow it with a variance of $\sigma^{2}_{x_{3R}}$. This can be
simply achieved by refining a series of test values of $\xi$ until the variance
of $a x_{2}^{\xi}$ matches $\sigma^{2}_{x_{3R}}$.

If both the density PDF and column density PDF are lognormal in form, the
appropriate value of $\xi$ can be simply derived. First, we note that 
equation~(\ref{eqno8}) is equivalent to:
\begin{equation}
\ln(x_{3R}) = \ln(a) + \xi \ln(x_{2}) ,
\label{eqno10}
\end{equation}
and therefore that the variances of $\ln(x_{3R})$ and $\ln(x_{2})$ are related by:
\begin{equation}
\sigma^{2}_{\ln(x_{3R})} =  \xi^{2} \sigma^{2}_{\ln(x_{2})} ,
\label{eqno11}
\end{equation}
independently of the value of $a$. 

With lognormal PDFs for density and column density:
\begin{equation}
\sigma^{2}_{\ln(x_{2})} =  \ln( 1 + \sigma^{2}_{x_{2}}) ,
\label{eqno12}
\end{equation}
and
\begin{equation}
\sigma^{2}_{\ln(x_{3R})} =  \ln( 1 + \sigma^{2}_{x_{3R}}) ,
\label{eqno13}
\end{equation}
so that:
\begin{equation}
\xi = \left ( \frac{\ln(1+\sigma^{2}_{x_{3R}})}{\ln(1+\sigma^{2}_{x_{2}})} \right )^{\frac{1}{2}} .
\label{eqno14}
\end{equation}
In general, the PDFs will not be lognormal in form and the more general procedure
for deriving $\xi$ should be employed.

An important component of the above reasoning (most clearly seen in 
equation~(\ref{eqno10}) which applies independently of the forms of the PDFs) 
is that we are assuming that
$\ln(x_{3R})$ is a scaled, shifted version of $\ln(x_{2})$, and therefore that the PDF of
$\ln(x_{3R})$ is a scaled, shifted version of the PDF of $\ln(x_{2})$. A test of whether the
3D normalised density PDF is recoverable by the above method is therefore also a test of whether
the {\it form} of the PDF of $\ln(x_{3})$ is preserved during the projection to 2D. Clearly this
will not be true in general but may be true of a restricted set of PDFs that characterise 
interstellar density fields. Some support for this is already in the literature (Ostriker et al
2001; V{\'a}zquez-Semadeni \& Garc{\'i}a 2001, Federrath et al 2009). In particular,
V{\'a}zquez-Semadeni \& Garc{\'i}a (2001) suggested that column density PDFs could appear
lognormal, but tended to appear normal if the correlation length of the
density field is small compared to the line-of-sight extent of the cloud. 
This is consistent with our picture: the correlation length is encoded in 
the power spectrum (e.g. by a turnover at some spatial frequency). The 
line-of-sight compression of the variance is controlled
by the power spectrum and is more pronounced for a flatter spectral slope. 
If the variance is reduced significantly, the density PDF can project into an 
(apparently) normal column density PDF, since a lognormal with very small variance is 
approximately normal. To constrain changes in the form of
the PDF under projection would require knowledge of how additional higher order moments
(beyond the second given by BFP) are related in 3D and 2D.

Finally, we note that scaling the PDF of $\ln(x_{3})$  to the PDF of $\ln(\rho) = \ln(x_{3}\rho_{0}) = \ln(x_{3}) + \ln(\rho_{0})$ is
trivial, as long as an estimate of $\rho_{0}$ is available. Lack of knowledge of $\rho_{0}$
does not affect the above analysis. However, it is important to recognise that a PDF
obtained at finite resolution may be a compressed version of the true PDF that would
be obtained in the continuous limit.

\section{Application to Numerical Simulations of Turbulence}

\begin{figure*}
\includegraphics[width=140mm]{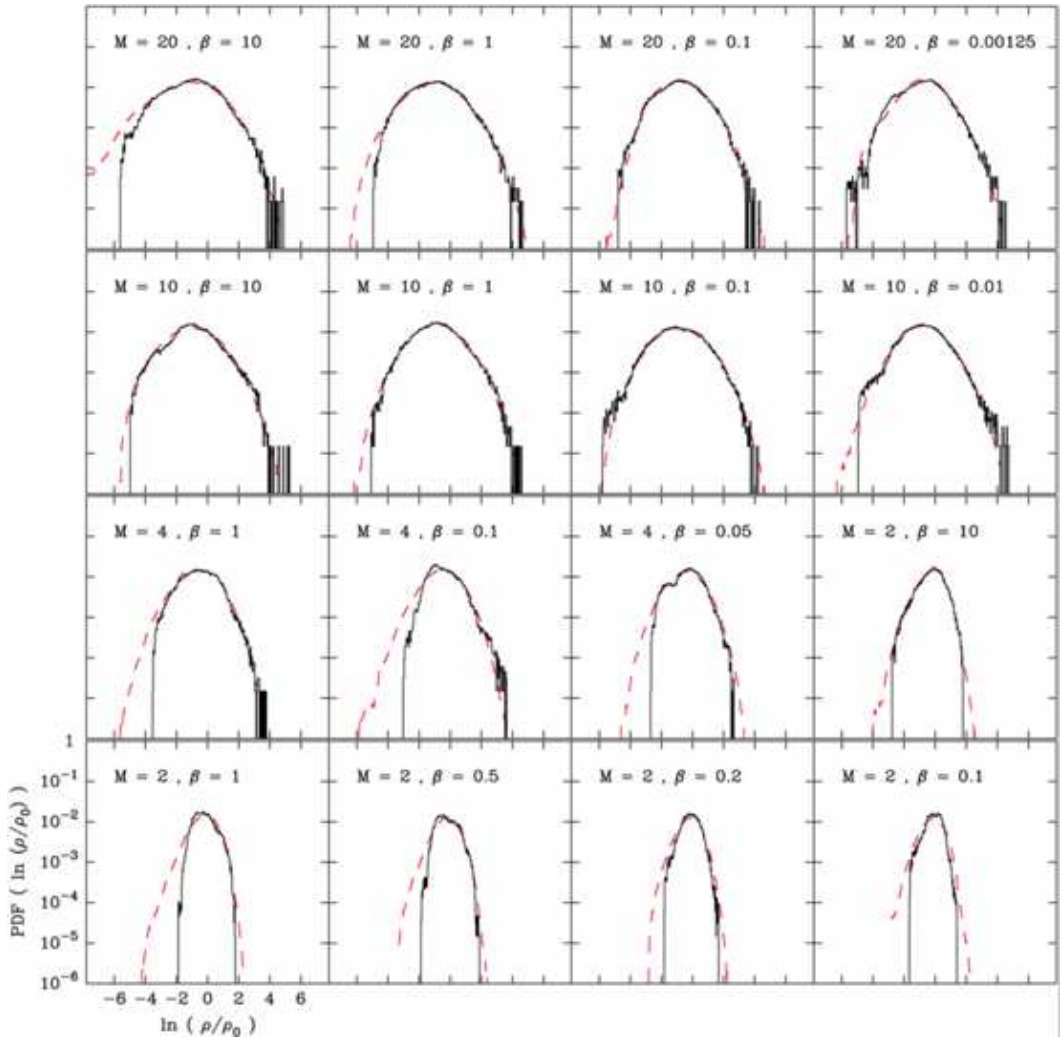}
  \caption
{ Comparison of true 3D PDFs of $\ln(x_{3})$ (dashed red lines) with reconstructed PDFs of $\ln(x_{3R})$
(black lines) generated from information contained solely in the projected column density field for the
MHD simulations. The parameters of each model (Mach number, $M$, and plasma $\beta$) are quoted in each panel.
}
\label{fig1}
\end{figure*}

\subsection{Magnetohydrodynamic Simulations}

We now test the method described above on numerically-simulated 
magnetohydrodynamic (MHD) turbulent
density fields, obtained at a range of Mach numbers ($M$~=~2~to~20) and
magnetic field strengths (ratio of thermal-to-magnetic pressure $\beta$~=0.00125~to~10).
The MHD simulations were computed using the grid-based code {\sc flash} (Fryxell et al 2000)
using solenoidal forcing of turbulence at large scales (Federrath et al 2009; Price \& Federrath 2010;
Brunt 2003; Brunt, Heyer, \& Mac Low 2009).
Previously, BFP have analysed these fields with the goal of determining the 3D 
normalised density variance using only 2D column density maps, so that values
of $R$ (equation~(\ref{eqno5})) are available for all fields. Here, we use a
representative subset of the simulated density fields to test the 3D PDF reconstruction method.

The basic procedure is as follows. Each 3D density field (scale ratio $\lambda = 256$) is 
integrated over one of the cardinal directions to produce a 2D column density field, which is
subsequently normalised to produce a field $x_{2}$, of mean unity. The normalised column
density variance, $\sigma^{2}_{x_{2}}$, is calculated for each field. The power spectrum
of $x_{2}$ is measured, and used to calculate $R$ via equation~(\ref{eqno5}), 
which then allows calculation of
$\sigma^{2}_{x_{3R}} = \sigma^{2}_{x_{2}} / R$. Since the PDFs of these fields are 
approximately lognormal, we calculated $\xi$ 
via equation~(\ref{eqno14}) and then transformed $x_{2}$ into $x_{3R}$ via equation~(\ref{eqno8}).
Finally, the PDF of $\ln(x_{3R})$ is calculated for comparison to the PDF of $\ln(x_{3})$ where $x_{3}$
is the true normalised 3D density field. The PDFs of $\ln(x_{3R})$ and $\ln(x_{3})$ are 
calculated over the same range and with the same bin width (256 bins).

In Figure~\ref{fig1} we show representative results obtained from MHD simulations
with a range of Mach numbers and values of the plasma $\beta$. We find that the PDF of 
$x_{3}$ is reconstructed with rather good accuracy, even though the forms of the PDFs are only
approximately lognormal, as we assumed in the calculation of $\xi$. The tails of the PDF (especially
the negative tails) are not well-recovered, but we note that $\lambda^{3}$ data points are included in
the calculation of the PDF of $\ln(x_{3})$ while only $\lambda^{2}$ data points are included 
in the calculation of the PDF of $\ln(x_{3R})$. Thus the extreme values of the field are less
likely to be present in $x_{3R}$ than in $x_{3}$, explaining the lack of sensitivity of the
reconstructed PDF to the tails of the true PDF. The worst PDF recovery is 
for the $M = 2$, $\beta = 0.1$ simulation, which contains the highest degree of anisotropy 
(caused by the strong magnetic field, which produces sub-Alfv{\'e}nic turbulence, as 
discussed in BFP). In general, the success of the PDF recovery is better at higher Mach number.

Taking into account the understood lack of sensitivity to the PDF tails, the success of the 
PDF reconstruction implies that the PDF of $\ln(x_{2})$ is, to a good approximation, a scaled, 
shifted copy of the PDF of $\ln(x_{3})$ -- i.e. that the log-space PDF retains its {\it form} 
under projection, but is compressed due to line-of-sight averaging. BFP provide a means of 
determining the amount of compression (via $R$ or, equivalently, $\xi$), which is dependent 
on the power spectrum of $x_{3}$, and measureable (under the assumption of isotropy) using 
the power spectrum of $x_{2}$.
We find that, for the simulations analysed here, $\xi$ is typically around 2.7, but varies by
about 0.5 around this value. Using an ensemble of slightly lower resolution numerical 
simulations ($\lambda = 128$), driven at a range of spatial scales, Brunt \& Mac Low (2004) 
found empirically that $\xi$~$\approx$~3. The $\lambda = 1024$ simulations of Federrath 
et al (2009) yield $\xi \approx 2.9$ and $\xi \approx 2$ for solenoidal and compressive 
forcing respectively. (In general, $\xi$ will depend on the scale ratio, $\lambda$, and the 
form of the power spectrum of the density field, and should be calculated in each application.)

The above results demonstrate that, to a good approximation, a lognormal density
PDF projects into a lognormal column density PDF. However, one may rightly question
the utility of the rescaling procedure. If we know the PDF is lognormal, and we
have an estimate of the variance, it is straightforward to generate an analytic
expression for the PDF. Indeed, we find that analytic lognormal functions scaled 
to a variance of $\sigma^{2}_{x_{3R}}$ are good representations of the density
PDFs displayed in Figure~\ref{fig1}. The true utility of the method must therefore
be demonstrated for density PDFs that deviate significantly from a lognormal
form. We conduct this test in the next section.

\subsection{Self-Gravitating Hydrodynamic Simulation}

\begin{figure*}
\begin{centering}
\includegraphics[width=174mm]{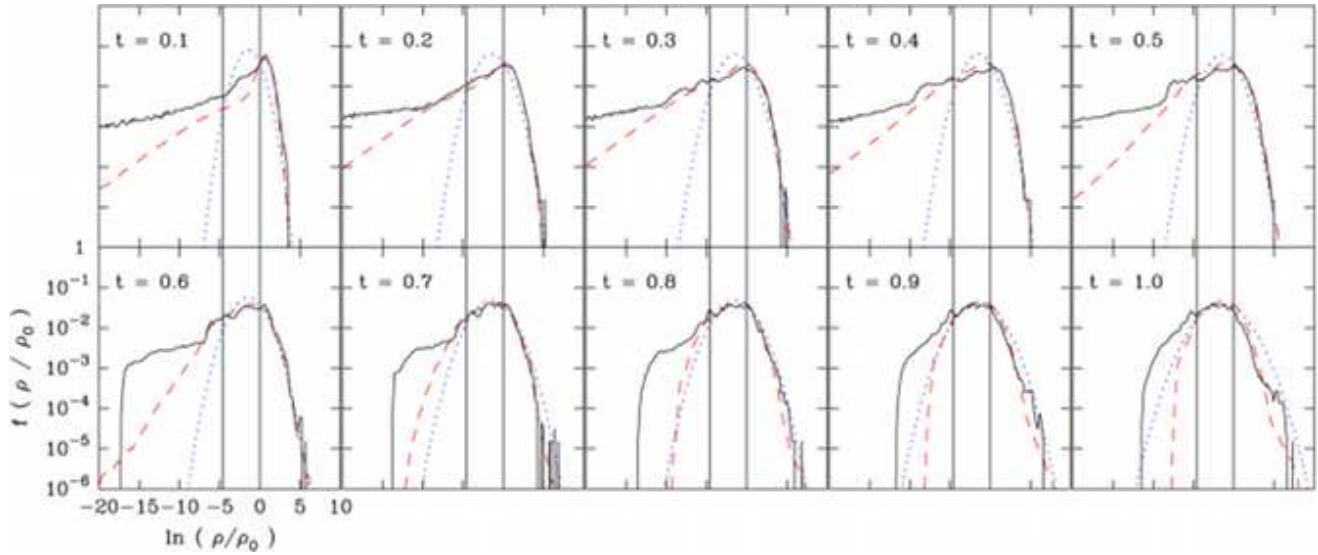}
  \caption
{ 
Comparison of reconstructed PDFs of $\ln(x_{3R})$ (black lines) with the true 3D PDFs of 
$\ln(x_{3})$ (red dashed lines) for the self-gravitating hydrodynamic simulation. The
time measured in free-fall times is shown in each panel. The
dotted blue lines show analytic lognormal functions with variance $\sigma^{2}_{x_{3R}}$ for
comparison. The vertical lines mark the mean density and one-hundredth of the mean
density in each PDF.
}
\label{fig2}
\end{centering}
\end{figure*}

We now test the method on a density field produced by a simulation of star formation in
self-gravitating hydrodynamic turbulence (Price \& Bate 2009). Density field snapshots 
were taken at 0.1, 0.2, ..., 0.9, 1.0~free-fall times. 
At the earliest times during the development of the turbulence and also at $t \sim t_{ff}$ once gravitational 
collapse has set in, the density PDFs differ substantially from a lognormal form and therefore
provide a good test of our method.
The projected column density fields were analysed using the BFP method and estimates of the 3D normalised
density variances were made (only one projection axis was used).
We rescaled each $x_{2}$ field by matching the variance
of $a x_{2}^{\xi}$ to $\sigma^{2}_{x_{3R}}$, where $\xi$ is found by a series of sequentially
refined test values. The resulting PDFs are shown in comparison to the true 3D PDFs 
in Figure~\ref{fig2}. In this figure, we also show analytic lognormal functions with
variance set to $\sigma^{2}_{x_{3R}}$ for each field. For reference, on each plot, 
we also draw vertical lines to mark the mean density and one-hundredth of the mean
density.

The evolution of the density PDF with time is as follows. At early times 
($t \lesssim 0.6~t_{ff}$) the PDF has an extended negative tail. A roughly 
lognormal form develops as $t$ approaches $t_{ff}$, accompanied by the
development of an extended positive tail.
Such extended positive tails are seen in column density PDFs and
are associated with gravitational collapse and star formation activity (Klessen 2000;
Federrath et al 2008; Kainulainen et al 2009). 

We find that, in general, the density PDF is recovered very well at densities above the
mean density. For most snapshots, the rescaled column density PDF is a reasonably good 
representation of the density PDF at densities as low as one-hundredth of the mean density, 
but deviates from it substantially below this. The extreme low density tail is difficult to measure
in most situations (especially in observational conditions which include noise) so this may not
be a significant issue. For all snapshots (except $t = 0.8~t_{ff}$) the rescaling
method performs better than using the analytic lognormal function. From the small number of
snapshots available, it appears that the positive tail seen at late times is relatively
more prominent in the rescaled column density field PDF than in the true density field PDF.
This suggests that observed positive tails in column density PDFs imply the 
presence of similar positive tails in density PDFs, but that some caution in their
interpretation should be applied.

The overall success of the method supports the idea that the form of the density PDF
is preserved in the column density PDF (even for a non-lognormal form) with 
appropriate provisos on the extreme positive and negative tails discussed above.

\section{Summary}

We have introduced and tested a simple method for reconstructing the probability density function (PDF)
of a 3D turbulent density field using information present solely in the projected (observable) column
density field in 2D. The method builds on a previously established method to calculate the 3D 
normalised density variance, recently presented by Brunt, Federrath, and Price (BFP, 2010). 

To a good approximation, the PDF of log(normalised column density) is a compressed, shifted version of the PDF of 
log(normalised density), but can deviate significantly in the extreme tails. 
The compression factor, $\xi$, can be derived observationally from the column density power 
spectrum, assuming statistical isotropy, using the BFP method.

\section*{Acknowledgments}
This work was supported by STFC Grant ST/F003277/1. We'd like to thank
the anonymous referee for good suggestions that improved the paper.
CB is supported by an RCUK fellowship at the University of Exeter, UK. 
CF is grateful for financial support by the International Max Planck
Research School for Astronomy and Cosmic Physics (IMPRS-A) and the
Heidelberg Graduate School of Fundamental Physics (HGSFP), which is
funded by the Excellence Initiative of the German Research Foundation
(DFG GSC 129/1). The \textsc{FLASH} MHD simulations were run at the
Leibniz-Rechenzentrum (grant pr32lo). The software used in
this work was in part developed by the DOE-supported ASC / Alliance
Center for Astrophysical Thermonuclear Flashes at the University of
Chicago. 

\hfill


\label{lastpage}

\end{document}